# 10.4 kW coherently combined ultrafast fiber laser

Michael Müller,[1,*] Christopher Aleshire,[1] Arno Klenke,[1,2] Elissa Haddad,[3] François Légaré,[3] Andreas Tünnermann,[1,2,4] and Jens Limpert [1,2,4]

[1]*Friedrich Schiller University Jena, Institute of Applied Physics, Albert-Einstein-Straße 15, 07745 Jena, Germany*
[2]*Helmholtz-Institute Jena, Fröbelstieg 3, 07743 Jena, Germany*
[3]*INRS, Centre Énergie Matériaux et Télécommunications, 1650 Blvd. Lionel-Boulet, Varennes, J3X1S2, Canada*
[4]*Fraunhofer Institute for Applied Optics and Precision Engineering, Albert-Einstein-Straße 7, 07745 Jena, Germany*
*\*Corresponding author: michael.mm.mueller@uni-jena.de*

**An ultrafast laser delivering 10.4 kW average output power based on coherent combination of twelve step-index fiber amplifiers is presented. The system emits close-to-transform-limited 254 fs pulses at 80 MHz repetition rate, has a high beam quality ($M^2 \leq 1.2$), and a low relative intensity noise of 0.56% in the frequency range of from 1 Hz to 1 MHz. Automated spatiotemporal alignment allows for hands-off operation.**

http://dx.doi.org/10.1364/OL.392843

High-power ultrafast lasers are demanded for industrial-scale precision materials processing, e.g. in solar cell and lithium battery production. The state of the art are ytterbium-doped thin-disk [1], slab [2] and fiber [3] chirped-pulse amplifiers delivering 1 kW-level average power in fundamental mode operation. Still, the heat load limits the achievable average power either by thermal lensing [4] in thin-disk and slab lasers or by transverse mode instability [5] in fiber lasers. Likewise, nonlinear refraction and Raman scattering set a limit to the attainable peak power.

Coherent beam combination (CBC, [6]) of several amplifiers in an interferometric setup allows to surpass these limits. In a typical scheme, *N* amplifiers are seeded by a common source and their output beams are made to interfere constructively, resulting ideally in a single output beam with up to *N*-times the brightness of a single amplifier channel. In practice, deviations in the spatial and temporal output characteristics of the amplifiers will result in a loss [7], which is quantified by the combining efficiency, defined as the ratio of combined power to the sum of all individual amplifier output powers. CBC allows for power scaling by orders of magnitude compared to a single amplifier, as the passive combination elements support higher peak and average power than the laser-active media. The price to pay is the need for phase stabilization to maintain constructive interference.

Fiber lasers are well suited for CBC due to their reproducible and simple setup. Numerous techniques for CBC of fiber amplifiers have been demonstrated, such as tiled-aperture combining based on diffraction [8] and filled-aperture combining using diffractive optical elements [9], dielectric polarization [10], and intensity beam splitters [11]. For Gaussian beams, filled-aperture combining is more efficient and hence preferred over tiled-aperture combining at theoretical maximum efficiencies of 100% and 68% [12]. In the filled-aperture schemes, intensity beam splitters offer some practical advantages, such as lower coating absorption compared to polarization beam splitters, lower cost than diffractive optical elements, and easy replacement in the case of damage, leading to the demonstration of a 3.5 kW average power ultrafast laser [11].

In this contribution, the successor of the system described in Ref. 11 is presented demonstrating improvement in all performance parameters, thus pushing the average power of ultrafast lasers to the 10 kW-level. A detailed description of the system is given, analysis of the output parameters is presented, and technical limitations as well as the further power scaling potential are discussed.

A schematic of the setup is depicted in Fig. 1. The seed source is an ultrafast fiber oscillator emitting pulses at 80 MHz repetition rate. The seed pulses are stretched in a set of chirped fiber Bragg gratings to 5 ns full width for the spectral hard cut of 14 nm centered at 1046 nm. Two core-pumped preamplifiers are embedded in this section to compensate for the grating transmission loss. A spectral amplitude and phase shaper is inserted to compensate for gain shaping and nonlinear phase accumulation in the later amplification stages, which improves the output pulse quality. Another two core-pumped preamplifiers raise the seed power from a few mW to 1 W at the end of this polarization-maintaining all-fiber front end.

The seed is free-space coupled into a 5 m long non-polarization-maintaining ytterbium-doped step-index fiber with a 20 µm core and a 400 µm cladding diameter (0.45 numerical aperture). The fiber is coiled to 12 cm diameter for higher order mode suppression, is equipped with plane-parallel antireflective-coated end caps, and is mounted in a water-cooled enclosure for robust high-power operation. Static polarization control via a quarter-wave plate (QWP) and a half-wave plate (HWP) results in stable linear output polarization. The amplifier is counter pumped with 250 W at 976 nm generating 150 W of seed power for the main amplifier stage. Up to this point, all amplifiers are optically isolated. Again in free space, the seed beam is split into 12 channels of equal power in a tree-type configuration of intensity beam splitters with 50% and 66% reflectivity. In all channels a set of QWP and HWP is employed for polarization control, and, with the exception of one reference channel, a piezo-actuated mirror (PA) is used for phase stabilization. The beams are coupled into 12 main amplifiers that are technically identical to the previous preamplifier, except that the active fibers are 11.00±0.02 m long and the pump power is up to 1.6 kW per channel, delivered from fiber-coupled non-wavelength-stabilized diodes. In free space after the amplifiers, two motorized mirrors (MM) and a motorized translation stage (MTS) are used to steer the beams into a second tree of beam splitters. The motorized components are remotely controlled for safe adjustment of the spatiotemporal overlap of the output beams. The combined beam emerges from a predefined port of the last beam splitter. All interference losses emerge from the 11 remaining open beam splitter ports in the combination stage and are terminated in water-cooled dumps. All optomechanical components starting from the beam combination stage are water-cooled to minimize thermal drift of the alignment due to cladding and stray light absorption. At the end of the beam combination stage, the beam diameter is increased from 3.3 mm to 6.5 mm $1/e^2$-diameter using a mirror telescope. Finally, the beam is sent through a double-pass Treacy-type compressor (8-fold diffraction) with 80% transmission efficiency based on dielectric gratings with a maximal length of 276 mm and 1740 lines per mm. Diffraction loss and depolarized light ejected from the gratings is terminated on water-cooled dumps. After the compressor, the leakage of a steering mirror is analyzed and the high-power beam is sent on a thermal power meter.

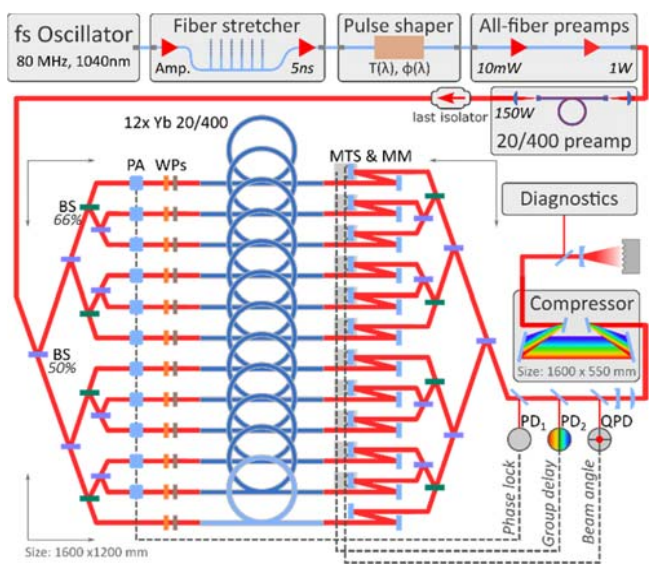

Fig. 1. Schematic setup of the laser system. BS: beam splitter, WP: quarter/half wave plate, MTS: motorized translation stage, MM: motorized mirror, (Q)PD: (quadrant) photodiode.

The interferometric superposition is susceptible to air currents and mechanical vibration, hence requires active stabilization. Here, locking of optical coherence by single-detector electronic frequency-tagging (LOCSET, [13]) is used. In this technique, small sinusoidal phase dithers at distinct frequencies are applied via the piezo-driven mirrors. This leads to small output intensity dithers, which are detected using a photodiode ($PD_1$). Demodulation of the photodiode signal with the initial sinusoids yields asymmetric error signals, featuring zero-crossings at the interference extrema of the respective channels. The error signals are used for closed-loop stabilization of the system at constructive interference via the piezo-driven mirrors. The dither frequencies are chosen starting at 6.5 kHz with 1.5 kHz spacing to 21.5 kHz to strike a balance between the requirements of having the dither frequencies higher than the typical phase perturbation frequencies, lower than the limit of the control electronics, and separated as far as possible for maximal stabilization bandwidth and minimal crosstalk.

Furthermore, an automated minimization of group delay [14] and pointing error [15] is implemented based on phase detection of the zeroth spectral and spatial interference fringes occurring upon misalignment. Here, LOCSET detection in off-centered spectral and spatial slices of the combined beam yields error signals telling the respective alignment errors during closed-loop phase lock. A short-pass filter transmitting half the optical spectrum to a photodiode ($PD_2$) forms the group delay detector. Horizontally and vertically joined sectors of a quadrant photodiode (QPD) make up the pointing mismatch detector. The error signal demodulation is implemented on a microcontroller. It calculates the delay, horizontal, and vertical pointing errors for the 11 dithered channels with respect to the ensemble using the original LOCSET phase dither and the photodiode signals. The alignment error signals are sent to a computer, which controls the motorized mirror mounts and linear translation stages. In an iterative procedure of error signal reading and mechanics adjustment, the spatial and temporal alignment of all channels converges to the reference channel, reaching maximum combining efficiency within a few minutes.

Prior to the combining experiment, all main amplifiers were tested at equal drive currents up to a pump power of ~1.6 kW, on average yielding 896±50 W output power after the compressor. The 6% standard deviation of the power is due to variations of the pump diode efficiency and the amplifier polarization purity. The resulting combination loss due to amplitude and nonlinear phase mismatch would be less than 1% [7].

For the initial beam combination, each amplifier was successively turned on, starting with the reference channel and sequentially adding adjacent channels. This was required, as the thermalization of the fibers going from passive to full-power operation imposed a relative optical path length drift of several hundred microns, measured by the emerging spectral interference fringes. This drift exceeds both the dynamic ranges of the piezo actuators (40 μm) and of the implemented group delay detection (200 μm). Hence, a onetime manual delay adjustment after about 5 minute warmup time was required for the 11 stabilized channels. Then, the automated alignment mechanism was used to optimize the interferometric overlap with respect to the already activated channels. This procedure was repeated for all channels and the output power was measured as is shown in Fig. 2. Finally, an output power of 10.4 kW was reached after compression, with a combination efficiency of 96%. At this operation point, the wall plug efficiency of the laser system is 20%. Once the path lengths were set in the thermalized system, it could be restarted by turning on all channels simultaneously. Then, after a single 5 minute thermalization time, the optical path lengths stabilized close enough to their last position such that the automated group delay detection could recover the optimum - essentially turn-key operation. The thermalized system worked without phase relock in excess of 30 minutes, as the path length drift stayed at about 3 µm peak-to-valley per minute, which is less than the piezo range.

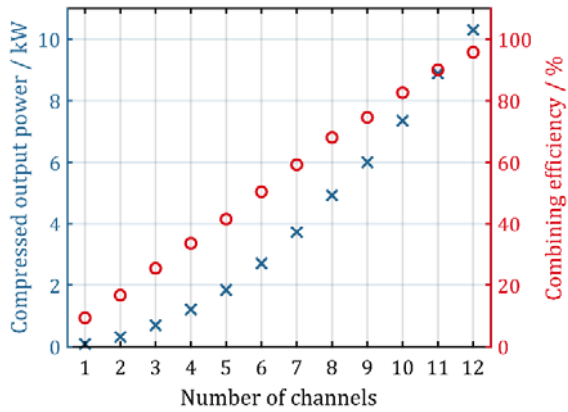

Fig. 2. Compressed output power and combining efficiency during sequential turn on of the amplifier channels.

The output beam quality was measured and it was found to be close-to-diffraction-limited with a maximal $M^2$-value of 1.2, as can be seen from the ISO-conform measurement shown in Fig. 3. The inset shows the corresponding beam profile of a well-confined Gaussian beam.

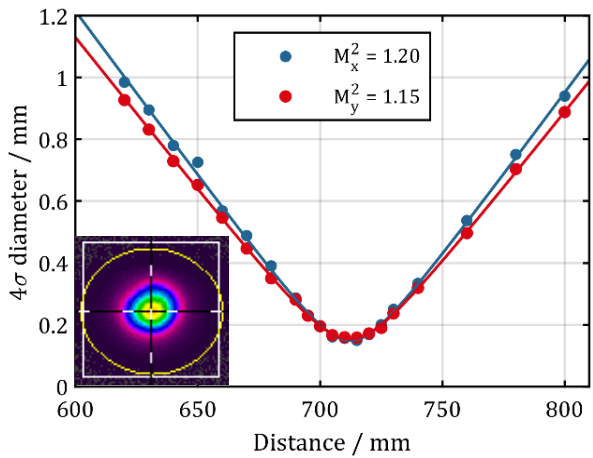

Fig. 3. ISO 11146-compliant $M^2$-measurement of the combined beam after the compressor with the 4σ-method. Inset: Beam profile.

In Fig. 4, optical spectra are shown for full power operation. A single amplifier barely shows the onset of stimulated Raman scattering at 1080 nm, indicating a low nonlinearity. Determined from simulation, the B-integral of the main amplifiers is approximately 5 rad. The amplified spontaneous emission below 1057 nm originates from the last preamplifier, as it does not attenuate upon combination unlike the longer wavelength components. The edge at 1057 nm itself is due to the dichroics in the amplifier units. After compression, the spectral shape of the pulses changes as the available dielectric gratings were optimized for 1035 nm leading to an overall reduced diffraction efficiency and higher loss for the long-wave part of the spectrum.

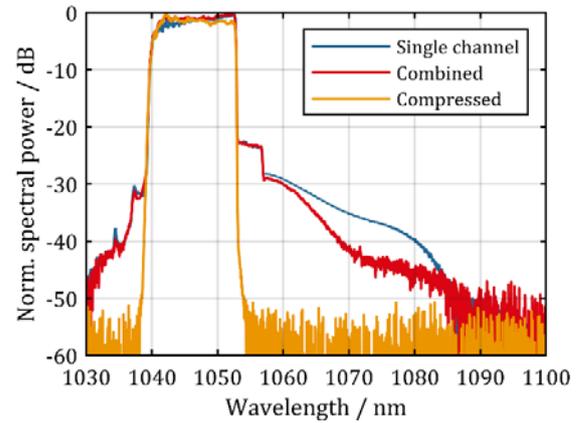

Fig. 4. Optical spectra at full power for a single channel before compression and for the combined beam before and after compression.

The pulse duration was characterized using non-collinear intensity autocorrelation, which is depicted in Fig. 5. For the combined beam, a FWHM pulse duration of 254 fs is measured, with a deconvolution factor of 1.33 determined from the spectral shape. The measured pulse duration is very close to the transform limit of 248 fs (calculated from Fig. 4). To achieve this result, the compression was optimized prior the combining experiment for a single amplifier channel at full power. First, a rectangular stretched pulse was shaped to minimize the nonlinearity. For that, a peak spectral amplitude suppression of 5 dB was sufficient as the short-wave gain in the all-fiber front end and the long-wave gain in the power amplifiers largely compensate each other. Then, the spectral phase was optimized using multiphoton interference phase scan [16].

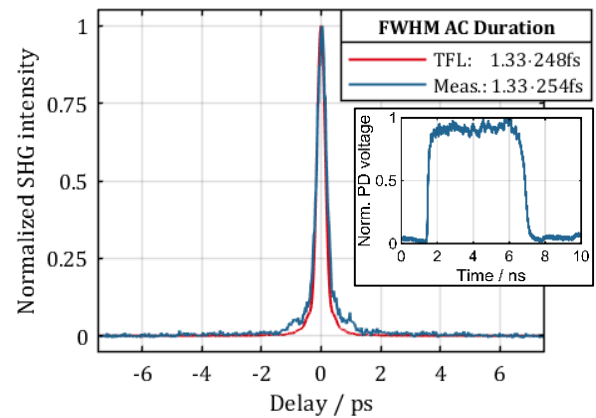

Fig. 5. Measured and calculated transform-limited (TFL) non-collinear intensity autocorrelation with deconvolution factor 1.33. Inset: Photodiode trace of the stretched pulse after the main amplifier.

Next, the power stability was measured, as shown in Fig. 6. A time series of the output intensity was acquired using an amplified photodiode, a 1 MHz low-pass filter and an oscilloscope in DC 50 Ω termination. Fourier analysis returns the power spectral density (PSD, top graph), which after integration yields the relative intensity noise (RIN, middle graph). The RIN of a single channel and of the combined beam at full power are very low, respectively at 0.08% and 0.56% in the frequency span from 1 Hz to 1 MHz. The combined beam noise has two almost equal contributions. First, the LOCSET dithering peaks and their harmonics between 6.5 kHz and 50 kHz and, second, a broad low-frequency contribution below 1 kHz. The LOCSET dithering is unavoidable, but its contribution could be reduced by shifting to higher dithering frequencies allowing for smaller dithering amplitudes. The low frequency contribution is a remainder of the system phase noise. It is linked to the cooling water flow and to the amplifier output power as it reduces with low power operation and in the absence of water cooling, but its origin in the components of the system is to be identified.

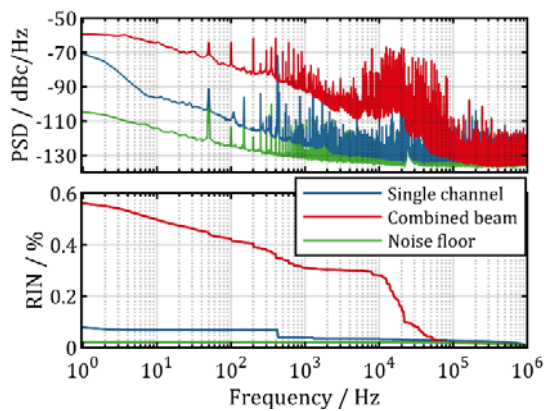

Fig. 6. Power spectral density (top) and integrated power spectral density (bottom) of a single channel and of the combined beam.

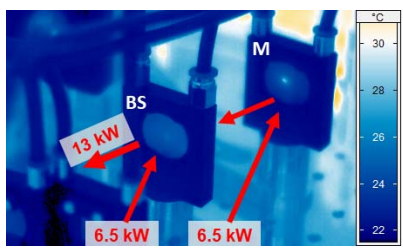

Fig. 7. Thermographic image of the last combination step overlaid with beam path and power. BS: final beam splitter, M: contaminated mirror.

Further power scaling ultimately will be limited by thermal lensing in the optical components. Thermographic imaging shown in Fig. 7 reveals that the final beam splitter heated up by only 3 K. Considering a temperature gradient of 30 K would still be tolerable on the beam splitter at the current beam diameter [17], the average power could be scaled to the 100 kW-level given more amplifier channels. At the same time, a steering mirror was observed to heat by 8 K, which, while still uncritical, is due to surface contamination in the laboratory environment and clean room assembly in a sealed housing will be needed in the future. In the compressor, no power-dependent beam quality degradation was observed, which indicates further scalability. Then, in contrast to the current system, the entire diffraction loss will have to be dumped outside the compressor chamber to avoid convection within.

In summary, an ultrafast laser system based on coherent beam combination of twelve ytterbium-doped step-index fiber amplifiers is presented. The system delivers 10.4 kW average power at a high combination efficiency of 96% with an excellent beam quality ($M^2 \leq 1.2$) and ultrashort pulses of 254 fs duration. The power noise is low at 0.56% RIN in the frequency range from 1 Hz to 1 MHz. The system features automated alignment, thus conveniently achieves and maintains high combining efficiency. The weak temperature gradient found on the final beam splitter allows to derive average power scalability of this technology to the 100 kW-level. The system as is could serve as research system for ultrafast materials processing. It demonstrates the feasibility of high-power beam combination systems as needed for future applications, e.g. in particle acceleration [18]. At the date of publication, it is the highest-average-power ultrafast laser.

**Funding.** Fraunhofer Cluster of Excellence Advanced photon sources (CAPS) and European Research Council MIMAS (670557)

**Disclosures.** The authors declare no conflicts of interest.